\documentclass[journal]{IEEEtran}

\ifCLASSINFOpdf
\else
   \usepackage[dvips]{graphicx}
\fi
\usepackage{url}
\usepackage{graphicx}

\usepackage[noadjust]{cite}
\usepackage{amsmath,amsthm,graphicx,cite,booktabs,amssymb,gensymb, color, xcolor, fancyhdr,float,perpage,tikz,gensymb, colortbl,enumerate,  braket}
\usepackage[linesnumbered,vlined,ruled]{algorithm2e}
\usepackage{algorithmic,array,enumerate,rotating}
\usepackage{subfigure,epsfig,dblfloatfix}
\usepackage{multirow,multicol}
\usepackage{hyperref}
\usetikzlibrary{matrix}
\hypersetup{
	colorlinks   = true,
	linkcolor    = red,
	citecolor    = red
}

\def\mcalbR{{\boldsymbol{\mathcal{R}}}}
\def\mcalbT{{\boldsymbol{\mathcal{T}}}}
\def\mbfOmega{{\boldsymbol{\mathbf{\Omega}}}}

\def\bsy{{\boldsymbol{y}}}
\def\bsH{{\boldsymbol{H}}}

\def\bsI{{\boldsymbol{I}}}

\def\bsC{{\boldsymbol{C}}}

\begin{document}

\title{Quantum Algorithm for Signal Denoising}

\author{Sayantan Dutta, \IEEEmembership{Member, IEEE}, Adrian Basarab, \IEEEmembership{Senior Member, IEEE},\\
Denis Kouam\'e, \IEEEmembership{Senior Member, IEEE}, Bertrand Georgeot 


\thanks{Corresponding author: Sayantan~Dutta (e-mail: sdu4004@med.cornell.edu; sayantan.dutta110@gmail.com).}

\thanks{S. Dutta is with the Department of Radiology, Weill Cornell Medicine, Cornell University, New York, New York, USA and the IRIT, Universit\'e de Toulouse, CNRS, Toulouse INP, UT3, Toulouse, France and the Laboratoire de Physique Th\'eorique, Universit\'e de Toulouse, CNRS, UPS, France.}

\thanks{A. Basarab is with the Universit\'e de Lyon, INSA‐Lyon, Université Claude Bernard Lyon 1, UJM-Saint Etienne, CNRS, Inserm, CREATIS UMR 5220, U1294, F‐69621, Villeurbanne, France.}

\thanks{D. Kouam\'e is with the IRIT, Universit\'e de Toulouse, CNRS, Toulouse INP, UT3, Toulouse, France.}

\thanks{B. Georgeot is with the Laboratoire de Physique Th\'eorique, Universit\'e de Toulouse, CNRS, UPS, France.}

\thanks{This work was supported by CNRS through the 80 prime program and NIH grant R01GM143388.}
\vspace{-8mm}
}

\markboth{Dutta \MakeLowercase{\textit{et al.}}: Quantum Algorithm for Signal Denoising}
{Dutta \MakeLowercase{\textit{et al.}}: Quantum Algorithm for Signal Denoising}
\maketitle

\begin{abstract}
This letter presents a novel \textit{quantum algorithm} for signal denoising, which performs a thresholding in the frequency domain through amplitude amplification and using an adaptive threshold determined by local mean values. The proposed algorithm is able to process \textit{both classical and quantum} signals. It is parametrically faster than previous classical and quantum denoising algorithms. Numerical results show that it is efficient at removing noise of both classical and quantum origin, significantly outperforming existing quantum algorithms in this respect, especially in the presence of quantum noise.
\end{abstract}


\IEEEpeerreviewmaketitle

\section{Introduction}

\IEEEPARstart{D}{enoising} a signal is an essential and long-standing challenge in number of applications, including medical science, geophysics, remote sensing, etc. Despite a large number of existing approaches, denoising is still the subject of active research nowadays.

Furthermore, with the development of quantum computing in recent years, such problems are bound to arise in a quantum setting as well. In this context, denoising is a different but complementary approach to error-correcting codes, which have been the subject of extensive research in the quantum computing community e.g, \cite{calderbank1996good, Ollivier2003Description, Xiong2022dual}. In particular, one could envision to use quantum computers to denoise a classical signal, as well as to deal directly with a quantum signal produced as the result of a quantum algorithm, or more generally as the output of a quantum process to be treated by quantum means.


With the advancement of quantum computing, many signal or image processing problems have been attempted to be solved using quantum algorithms in the last decade, with promising performance achieved in image compression \cite{latorre2005image}, encryption \cite{liu2019quantum}, watermarking \cite{song2014dynamic, iliyasu2012watermarking, zhang2013watermark}, color translation \cite{jiang2015quantum}, resolution enhancement \cite{zhao2022dual}, feature extraction \cite{zhang2015local}, edge detection \cite{yao2017quantum, peng2019automated}, quantum image representation \cite{le2011flexible, zhang2013neqr}, quantum machine learning \cite{nikoloska2022training}, and so on.
Quantum denoising algorithms for removing standard classical noise (\textit{e.g.}, Gaussian, salt \& pepper, etc.) have also emerged in recent years. In particular, spatial filtering methods such as quantum median filtering \cite{li2018quantum, jiang2019improved} or weighted average filtering \cite{li2017quantum,yuan2017quantum} have been reported in the literature. Likewise, transform domain thresholding methods based on Quantum Fourier Transforms (QFT) \cite{caraiman2013quantum, li2018improved} and Quantum Wavelet Transforms (QWT) \cite{chakraborty2020image} have also been proposed.

However, in the literature on classical denoising, it has been seen that such classical transform methods are usually less powerful and efficient than methods which use tailored transforms adapted to the properties of the signal to be processed. Among recent works, we have developed a method based on adaptive transforms built from solutions of the Schroedinger equation \cite{dutta2021quantum}. The main advantage of such methods is that the thresholding is automatically modified according to the local properties of the underlying signal. Direct implementation of the method of \cite{dutta2021quantum} on a quantum computer or device is not an easy problem. Therefore, in the present work, we propose a modified algorithm which still takes advantage of the local properties of the signal while being efficiently implementable on a quantum computer. The main idea is to locally threshold a noisy quantum state (in quantum computation, a signal is represented by a quantum state) in the frequency domain, using combinations of QFT and amplitude amplification, a process derived from Grover's quantum search algorithm \cite{grover1996fast, brassard2002quantum}. We show how to apply our method to a classical signal transformed into a quantum state by a specific process, as well as to a quantum state produced by a purely quantum process such as a quantum algorithm. We evaluate the complexity of our algorithm and show its efficiency compared to previous algorithms. Finally, we present numerical results demonstrating that our algorithm can denoise both classical and quantum signals, and more efficiently than previously proposed algorithms.

\vspace{-2mm}
\section{Quantum Denoising Algorithm}
\label{sec:quan_algo}


Our algorithm can be applied to denoise an initially noisy quantum state on a N-dimensional Hilbert space, which can be the end-product of a quantum algorithm running on a quantum computer such as algorithms simulating quantum physical systems. It can also apply to a classical numerical signal $\bsy \in \mathbb{R}^N$. In the latter case, the classical signal should be first implemented in a quantum state. 
In both cases, to use our denoising algorithm, we will need to additionally store in a quantum register the local mean value of the signal in sliding windows, as well as the value of a local threshold. To learn more about the standard tools of quantum computing, we refer the reader to good introductions to the field such as  \cite{steane1998quantum, nielsen2010quantum}.

In the case of a noisy classical signal $\bsy \in \mathbb{R}^N$, for simplicity and without loss of generality, we consider $N$ as a power of two such that $N=2^n$. Our procedure requires to divide $\bsy$ into $P=2^p$ signal windows of size $M=2^m$, such that $N=PM$. In order to store the values of $\bsy$ we need a $n$-qubit register split into a $m$-qubit register denoted $\ket{i}^{\mbox{\scriptsize{I}}}$ which stores the position inside a signal segment and a $p$-qubit register denoted $\ket{j}^{\mbox{\scriptsize{J}}}$ which holds the segment number. We also need two additional registers initially in the state $\ket{0}$. The first one of $a$ qubits will hold the mean values inside a segment as a binary sequence for each segment number. The second one of $b$ qubits will hold the local threshold which will be used for this segment. 
The mean value of a local window $A_j$ should be precomputed classically, as well as the thresholding function $\tau(A_j)$ which depends on the mean value $A_j$, giving a sample-dependent adaptive threshold; assigning precise values to the function $\tau(A_j)$ can be done classically before running the algorithm following methods in the line of the one explained in \cite{dutta2022novel}.
$\bsy$ is encoded into a quantum state which contains at the same time the values of the signal as amplitudes and its sliding windows mean values in a register, using a modification of a known method (see e.g. \cite{le2011flexible, zhang2013neqr}) that we detail below.


\subsubsection{Quantum Model}
\label{sec:quan_model}

The preparation of the quantum state starts by initializing the registers to $\ket{0}$, thus leading to the state $\ket{0}  \ket{0}^{\mbox{\scriptsize{I}}}  \ket{0}^{\mbox{\scriptsize{Mean}}} \ket{0}^{\mbox{\scriptsize{J}}}  \ket{0}^{\mbox{\scriptsize{Thre}}}$, where registers `I',  `Mean', `J', and `Thre', respectively, contain $m$-qubits for the computational basis, $a$-qubits for the mean values, $p$-qubits for the segment number, and $b$-qubit for the local threshold.
Single-qubit Hadamard gates $\bsH = \frac{1}{\sqrt{2}} \big( ( \ket{0} + \ket{1} ) \bra{0}  + ( \ket{0} - \ket{1} ) \bra{1}  \big)$ are applied on the registers $\ket{0}^{\mbox{\scriptsize{I}}}$ and $\ket{0}^{\mbox{\scriptsize{J}}}$ to get an intermediate superposition state
\begin{equation}
\ket{\psi}_1 = \frac{1}{\sqrt{PM}} \sum_{j=0}^{P-1} \sum_{i=0}^{M-1} \ket{0} \ket{i}^{\mbox{\scriptsize{I}}} \ket{0}^{\mbox{\scriptsize{Mean}}} \ket{j}^{\mbox{\scriptsize{J}}} \ket{0}^{\mbox{\scriptsize{Thre}}},
\label{eq:state_1}
\end{equation}
with all possible superposition states, where $ \sum_{i=0}^{M-1} \ket{i}^{\mbox{\scriptsize{I}}} $ and $ \sum_{j=0}^{P-1} \ket{j}^{\mbox{\scriptsize{J}}} $ represent the superpositions of the computational basis states and segment numbers, respectively. Hence, for any two segments, their respective computational basis states will be orthogonal to each other.

Classical digital signals typically are encoded in quantum states using amplitude representation method \cite{le2011flexible}, where signal values are stored in the amplitudes of the qubits, or using basis representation method \cite{zhang2013neqr}, where the basis state of a qubit sequence register the signal values and positions. 
To encode the signal values $s_{ji}$ (where, $s_{ji}$ is the $i$-th signal value of the $j$-th signal segment) to the respective computational basis states $\ket{i}^{\mbox{\scriptsize{I}}}$, the mean value $A_j$ and the threshold $\tau(A_j)$ for each state $\ket{j}^{\mbox{\scriptsize{J}}}$, we combine these two techniques and define a unitary transform $\mcalbT$,
using $\mcalbR_y(\theta_{ji})$, controlled rotation operations about the $y$-axis with angle $\theta_{ji}$. These angles $\theta_{ji} = \frac{\pi}{2^a} s_{ji}$ encode the information of the signal in the amplitudes of the quantum state \cite{le2011flexible}.
We combine those with quantum operations $\mbfOmega_j$ and $\bsC_{\mbfOmega_j}$, being two controlled value setting operations that encode the mean value $A_j$ to the register $\ket{0}^{\mbox{\scriptsize{Mean}}}$ and the threshold $\tau(A_j)$ to the register $\ket{0}^{\mbox{\scriptsize{Thre}}}$, respectively for each segment. The unitary operation $\mcalbT$ is explicitly laid out in the Supp. Mat.

Through the unitary operation $\mcalbT$, the state $\ket{\psi}_1$ becomes
\begin{align}
\ket{\psi}_2 = 
& \frac{1}{\sqrt{PM}} \sum_{j=0}^{P-1} \sum_{i=0}^{M-1} \Big( \mbox{cos}(\theta_{ji})\ket{0}   \nonumber\\
& + \mbox{sin}(\theta_{ji})\ket{1} \Big) \ket{i}^{\mbox{\scriptsize{I}}} \ket{A_j}^{\mbox{\scriptsize{Mean}}} \ket{j}^{\mbox{\scriptsize{J}}} \ket{\tau(A_j)}^{\mbox{\scriptsize{Thre}}},
\label{eq:state_cos}
\end{align}
where both $\mbox{cos}(\theta_{ji})$ and $\mbox{sin}(\theta_{ji})$ store the same signal information. Therefore, to obtain a more simplified state we make a measurement of the first register, collapsing the state $\ket{\psi}_2$ into the desired state $\ket{\psi}$ given by eq.~\eqref{eq:state_psi}, where $ \alpha_{ji} = \mbox{cos}(\theta_{ji})$ preserves the noisy signal details of the classical signal $\bsy$ in our proposed quantum representation.
Thus the prepared noisy quantum state $\ket{\psi}$ is given by
\begin{equation}
\ket{\psi}= \sum_{j=0}^{P-1} \sum_{i=0}^{M-1}
\alpha_{ji} \ket{i}^{\mbox{\scriptsize{I}}} \ket{A_j}^{\mbox{\scriptsize{Mean}}} \ket{j}^{\mbox{\scriptsize{J}}} \ket{\tau(A_j)}^{\mbox{\scriptsize{Thre}}},
\label{eq:state_psi}
\end{equation}
where the noisy signal details are encoded in the amplitudes $\alpha_{ji}$ of the computational basis states $\ket{i}^{\mbox{\scriptsize{I}}}$. As mentioned above, the registers $\ket{A_j}^{\mbox{\scriptsize{Mean}}}$ and $\ket{\tau(A_j)}^{\mbox{\scriptsize{Thre}}}$ store the mean value $A_j$ and threshold $\tau(A_j)$ respectively for each segment.

In the case where one starts directly from a quantum signal, it will be initially in the form:
\begin{equation}
\ket{\psi_0}=   \sum_{i=0}^{N-1}
\alpha_{i} \ket{i}, 
\label{eq:state_psi0}
\end{equation}
where the noisy signal details are encoded in the amplitudes $\alpha_{i}$ of the computational basis states $\ket{i}$.
One should also transform it into a state of the form \eqref{eq:state_psi}. If the state \eqref{eq:state_psi0} is the end product of a quantum algorithm, this requires to run the algorithm a certain number of times, of the order of $P$ times, with a measurement of the first $p$ qubits at the end. This will allow to get an estimate of $A_j$ in each segment. The quantum noise will not be necessarily  the same for each run of the algorithm, but this is not a problem, since the algorithm does not need a very accurate value for $A_j$. Then one can run the algorithm one more time, and use the quantum operations $\mbfOmega_j$ and $\bsC_{\mbfOmega_j}$ above to produce the state \eqref{eq:state_psi}. One can process similarly if the state \eqref{eq:state_psi0} is produced through a quantum process other than a quantum algorithm, the only requirement being that the process can be repeated a certain number of times with results that are similar enough to each other to construct a meaningful average mean value for each segment.

\subsubsection{Denoising Scheme}
\label{sec:denoi_scheme}

Once the state \eqref{eq:state_psi} is built, the implementation of the proposed denoising quantum algorithm uses a $m$-qubits QFT implemented on the register $\ket{i}^{\mbox{\scriptsize{I}}}$ to transform the computational basis states into the frequency basis states $\ket{k}^{\mbox{\scriptsize{I}}}$, which gives
\begin{equation}
\ket{\xi}= \sum_{j=0}^{P-1} \sum_{i=0}^{M-1} \beta_{ji} \ket{k}^{\mbox{\scriptsize{I}}} \ket{A_j}^{\mbox{\scriptsize{Mean}}} \ket{j}^{\mbox{\scriptsize{J}}} \ket{\tau(A_j)}^{\mbox{\scriptsize{Thre}}},
\label{eq:state_xi_qft}
\end{equation}
where 
$\beta_{ji}$ are the Fourier coefficients.

The denoising is processed in the frequency domain using an adaptive thresholding which depends on the mean-value. High frequency components are generally dominated by the noise, which is a primary hypothesis in signal processing. Our algorithm uses a local threshold specific to each segment to select the most pertinent low-frequency components of the amplitudes in the Fourier basis. In order to select efficiently these components, we use amplitude amplification, based on
Grover's quantum search algorithm \cite{grover1996fast, brassard2002quantum}, which enables to amplify the amplitudes of some marked states, which are low-frequency states in our problem, while consequentially reducing that of the unmarked states, which are higher in frequency than the threshold in our case. The Grover's algorithm consists of two operators: oracle and diffusion. The oracle marks those basis states $\ket{k}^{\mbox{\scriptsize{I}}}$ that satisfy the thresholding conditions $ k - \tau(A_j)  \leq 0$ with a negative phase by rotating the phase by $\pi$ radians.\footnote{The operation of the oracle on an arbitrary state $\ket{w}$ can be expressed as $\ket{w} \xrightarrow{\mbox{oracle}} (-1)^{f(w)} \ket{w}$, where $f(w) = 1$ if $w$ satisfy the thresholding condition and $f(w) = 0$ otherwise.}
Thus the oracle helps to identify the low-frequency states without actually measuring them. The action of the diffusion operation can be interpreted as successive rotations of the marked state to bring it closer to the subspace containing the solutions of the search problem \cite{vinod2021finding}.
After $O(\sqrt{M/\sum_{j=0}^{P-1} M_j})$ repetitions of these two operations, where $M_j$ is the number of marked states in the $j$-th signal-sample (note that $M_j \leq M$ for all $j$), the amplitudes of the marked basis states $\ket{k}^{\mbox{\scriptsize{I}}}$ in each segment are amplified and the ones of the unmarked states decrease. Therefore, the proposed thresholding process first marks the low-frequency states depending on the threshold value $\tau(A_j)$ for each segment, followed by Grover's algorithm to threshold the high-frequency contributions associated with the noise, in a way adapted to the local mean of the signal in each segment. This process leads to a modified state $\ket{\phi}$ given by
\begin{equation}
\ket{\phi}= \sum_{j=0}^{P-1} \sum_{i=0}^{M-1} \gamma_{ji} \ket{k}^{\mbox{\scriptsize{I}}} \ket{A_j}^{\mbox{\scriptsize{Mean}}} \ket{j}^{\mbox{\scriptsize{J}}} \ket{\tau(A_j)}^{\mbox{\scriptsize{Thre}}},
\label{eq:state_phi_grover}
\end{equation}
where 
$\gamma_{ji}$ are the thresholded amplitudes obtained by performing Grover's amplitude amplification.

Finally, by applying $m$-qubits inverse-QFT (IQFT) on the register $\ket{k}^{\mbox{\scriptsize{I}}}$ of the state $\ket{\phi}$ in eq.~\eqref{eq:state_phi_grover} one returns to the computational basis states $\ket{i}^{\mbox{\scriptsize{I}}}$, and discarding the registers holding $A_j$ and $\tau(A_j)$ which are no longer needed, one gets
\begin{equation}
\ket{\varphi}= \sum_{j=0}^{P-1} \sum_{i=0}^{M-1} \delta_{ji} \ket{i}^{\mbox{\scriptsize{I}}}  \ket{j}^{\mbox{\scriptsize{J}}},
\label{eq:state_varphi_recov}
\end{equation}
where 
$\delta_{ji}$ are the thresholded amplitudes associated with the denoised signal.


\section{Complexity of the algorithm}
\label{sec:time_space_complex}

Starting from the quantum signal \eqref{eq:state_psi0}, the gate operations needed to perform the proposed algorithm  correspond to the QFT, the amplitude amplification algorithm, and finally the IQFT. The $m$-qubits QFT and IQFT require $O(\mbox{log}^2 M)$ operations, and the amplitude amplification algorithm needs $O(\sqrt{N/\sum_j M_j})$, where $M_j$ is the number of states below threshold in the $j$-th segment (with $M_j \leq M$ for all $j$).
Classically the whole process starting from the same premises would be dominated by the Fourier transforms and would cost $O(PM\mbox{log}M)$ operations, \textit{i.e.}, $P$ Fourier transforms should be performed, for each window of size $M$. The gain of the proposed algorithm compared to the classical one depends on the size of the patches $M$ and the number of states selected by the threshold. 
The smaller the patch, the more efficient becomes the proposed algorithm; at worse, for extremely low threshold, the amplitude amplification will cost $O(\sqrt{N})$ and will dominate the computational cost, with still a quadratic gain compared to classical algorithms. For large or intermediate threshold, the QFT dominates and the cost is $O(\mbox{log}^2 M)$ compared to $O(PM\mbox{log}M)$.
Furthermore, $O(\mbox{log}N)$ qubits of storage space is sufficient for our proposed quantum computation in contrast with the classical process with $O(N)$ bits storage. 
To be fair, it should be noted that in the case of a purely classical signal, this comparison is done without counting the cost of transforming the classical signal into the quantum initial signal \eqref{eq:state_psi0}, which using the procedures of \cite{le2011flexible, zhang2013neqr} would cost $O(N^2)$ operations; however using more refined implementations of the controlled operations \cite{liu2007analytic}, this could be reduced to $O(N\mbox{log}^2 N)$ operations. 

We also note that compared to other quantum algorithms for denoising based on simple transforms and direct thresholding (e.g. through measurements) the proposed algorithm leads to a gain on two distinct parts of the algorithm; indeed, previous quantum algorithms using QFT \cite{caraiman2013quantum} or QWT \cite{chakraborty2020image} require $O(\mbox{log}^2 N)$ operations for the transform and 
$O(N/\sum_j M_j)$ operations for the thresholding. For $M\ll N$ or $\sum_j M_j\ll N$, these two parts are significantly accelerated by our algorithm.


\begin{figure*}[t!]
\centering

\subfigure[Denoising performance in the presence of conventional additive white Gaussian noise.]
{\includegraphics[width=1\textwidth]{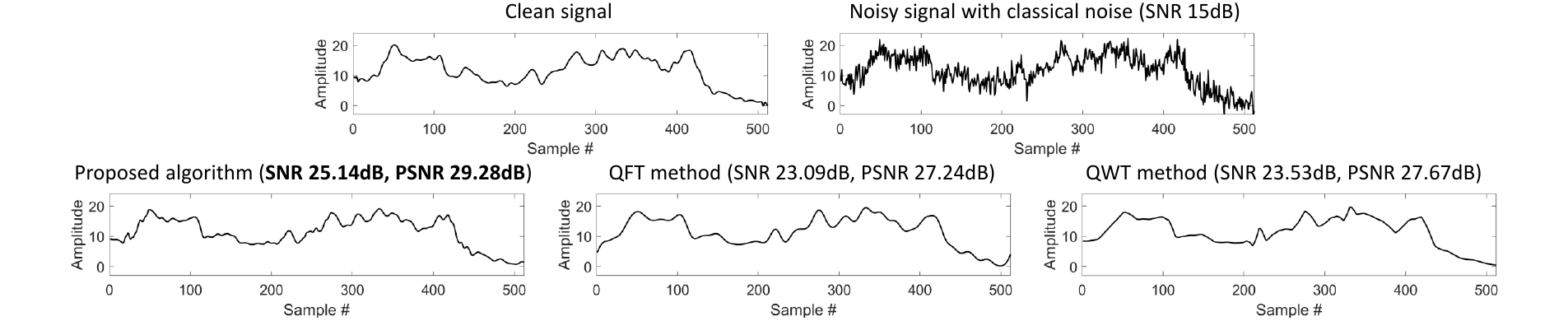}}

\subfigure[Denoising performance in the presence of quantum phase noise generated due to random angles in the phase of basic quantum gates.]
{\includegraphics[width=1\textwidth]{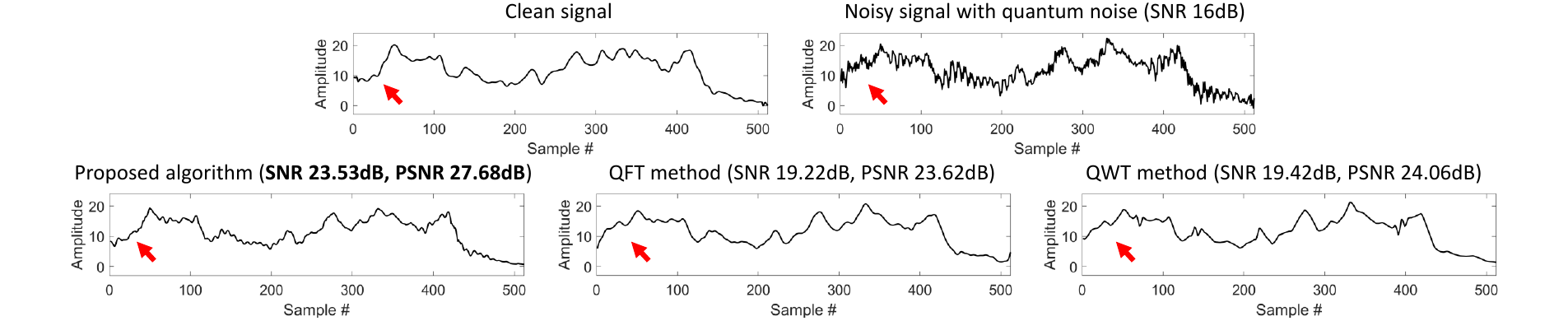}}

\subfigure[Denoising performance in the presence of mixed noise (classical Gaussian + quantum phase noise).]
{\includegraphics[width=1\textwidth]{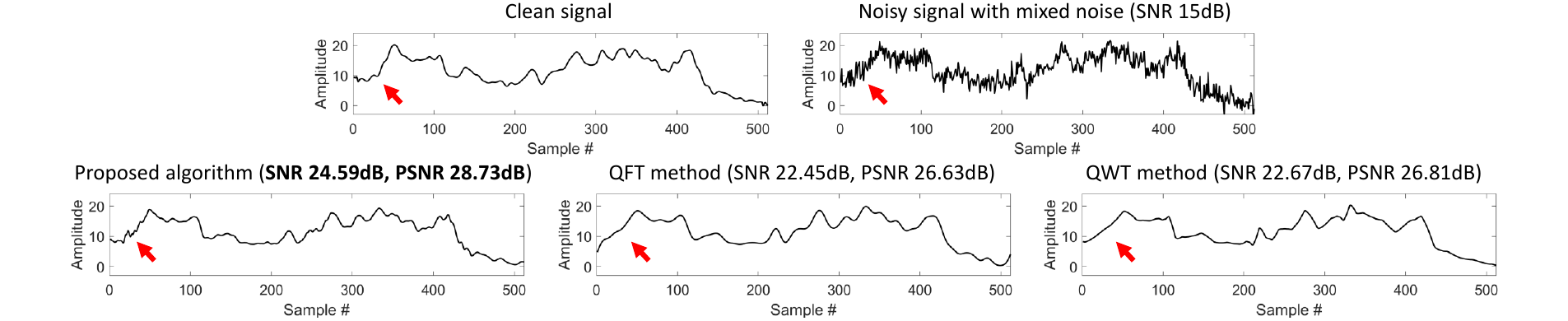}}

\caption{Denoising potential of different methods in various noise scenarios. The best results are highlighted in bold. The red arrow shows the reconstruction distortion arising in the presence of quantum noise while using QFT and QWT, where the proposed method efficiently preserves the shape.}
\label{fig:denoi_result}
\vspace{-5mm}
\end{figure*}

\section{Numerical Results}
\label{sec:results}

In this section, we assess the performance of our proposed quantum denoising algorithm in presence of: (i) classical noise, (ii) quantum phase noise, and (iii) mixed classical and quantum noise. The simulations were executed on a classical computer using MATLAB implantation based on linear algebra, where quantum states and the unitary transformations are represented by complex vectors and unitary matrices respectively.

(i) Classical noise: Conventional additive white Gaussian noise (AWGN) with signal-to-noise-ratio (SNR) 15dB is considered as the classical noise that contaminates the classical signal. Following the quantum signal representation scheme proposed in Sec.~\ref{sec:quan_algo}, the noisy quantum state $\ket{\psi}$ is prepared before implementing our proposed quantum algorithm. Fig.~\ref{fig:denoi_result}(a) depicts the respective denoising performance.

(ii) Quantum phase noise: Unitary transformations in a quantum computer are performed by some quantum circuits, basically made of one-qubit or two-qubit unitary operations or gates. To implement this kind of noise, which can correspond to, e.g., noise in the experimental implementation of the gates, we perform a QFT followed by an IQFT, and rotate each of the basic unitary operations with a small random angle of amplitude $\epsilon = 10^{-1}$.  These random $\epsilon$ rotations vary in time, generating a phase noise in the unitary transformations \cite{song2001quantum, georgeot2001exponential}.
The circuit implementation of our proposed quantum denoising algorithm is then implemented free of noise. The denoising performance in presence of such quantum noise is shown in Fig.~\ref{fig:denoi_result}(b).

(iii) Mixed noise: Finally, AWGN and quantum phase noise are combined to generate the effect of a mixed noise. A classical noisy signal contaminated with AWGN is used to prepare the quantum state $\ket{\psi}$ in eq.~\eqref{eq:state_psi} following the strategy proposed in Sec.~\ref{sec:quan_algo} and then  quantum phase noise $\epsilon = 10^{-1}$ is added as in (ii). Fig.~\ref{fig:denoi_result}(c) illustrates the denoising ability of our proposed algorithm in presence of mixed noise.

Rigorous comparisons\footnote{In the context of classical signal denoising, a vast literature with sophisticated algorithms is available, e.g., \cite{van2009sure, zuo2016image, yang2018bm3dnet, zhu2019seismic, dutta2022deep, dutta2022diva}. Since our main focus in this letter is quantum computation, we only consider for comparison algorithms that have been proposed for a quantum computer.} with existing quantum algorithms based on QFT \cite{caraiman2013quantum} and QWT \cite{chakraborty2020image} are reported under the three noise scenarios. Visual and quantitative results in terms of SNR and peak-SNR (PSNR) in Fig.~\ref{fig:denoi_result} indicate that the proposed quantum algorithm significantly outperforms other methods in all scenarios. We observe that in presence of quantum phase noise and mixed noise the proposed quantum denoising algorithm respectively presents a gain of more than 3.5 dB and 2 dB PSNRs compared to the second best performing method using QWT. This is particularly visible on the extreme left of the signal, where the proposed algorithm reproduces the global shape which is lost in the denoising by QFT and QWT alone.
Our results thus indicate that simpler QFT and QWT based algorithms are less efficient, and in particular less adapted to the quantum phase noise than the proposed one; a result we attribute to the locally adaptive nature of our method.
Other types of noise are possible, and in particular the quantum signal and quantum registers can be subject to decoherence. In order to test our algorithm in presence of another type of quantum noise which appears in decoherence processes, we  added to the process bit-flip errors on the registers. On average, our proposed method produces denoised outputs with 19.77 dB PSNR, while QFT and QWT algorithms yield 17.98 dB and 17.46 dB PSNRs respectively (more details and an explicit example can be found in the Supp. Mat.). This suggests that our method is efficient in presence  of other quantum noise and in particular quantum decoherence noises.
Denoising results on classical Poisson and mixed (Poisson + quantum) noise scenarios are also reported in the Supp. Mat., showing that our method is equally effective in the presence of signal dependent noise models such as Poissonian noise.






\section{Conclusion}
\label{sec:conclusion}

This letter introduced a novel quantum algorithm for signal denoising. Our proposed algorithm uses a local thresholding in the frequency domain depending on the local mean value of the signal, which provides an adaptive way of thresholding a noisy signal.
To illustrate the feasibility of our proposed method, we conducted experiments in presence of classical Gaussian noise, quantum phase noise in quantum gates, bit-flip noise and mixed noises (combined Gaussian and quantum phase noise).
Our results suggest that our quantum algorithm is able to provide better denoising performances than the previously proposed quantum algorithms, especially in the presence of quantum phase noise, while being parametrically faster than existing classical and quantum algorithms. We note that such denoising performed a posteriori may be important to harness noisy quantum computers, and can be substituted or combined with standard quantum error correction, performed during the quantum computation. It should be noted that due to the efficiency of the algorithm and its ability to self-correct the noise produced, the signal size that could be handled should grow exponentially with the inverse of the level of noise deemed acceptable.
In this letter, we empirically choose a simple thresholding operation based on the mean value, whereas the proposed method provides a general thresholding framework. Thus it would be interesting to study other advanced thresholding schemes, such as Wiener filtering incorporating mean and variance in the thresholding process.
A promising future work could also be to extend our algorithm to other imaging applications, such as deconvolution.





\bibliographystyle{IEEEbib}
\bibliography{Quantum_Denoising_Algoeithm}


\onecolumn
\section*{\large{Supplementary Material:}\\ \LARGE{Quantum Algorithm for Signal Denoising}}
\label{sec:supplement}

\begin{center}

Sayantan Dutta\footnote{Corresponding author: Sayantan~Dutta (e-mail: sdu4004@med.cornell.edu; sayantan.dutta110@gmail.com)}, Adrian Basarab, Denis Kouam\'e, and Bertrand Georgeot 

\end{center}

\maketitle



\begin{figure*}[h!]
\centering
\vspace{-2mm}
\subfigure[Denoising performance in the presence of conventional Poisson noise.]
{\includegraphics[width=1\textwidth]{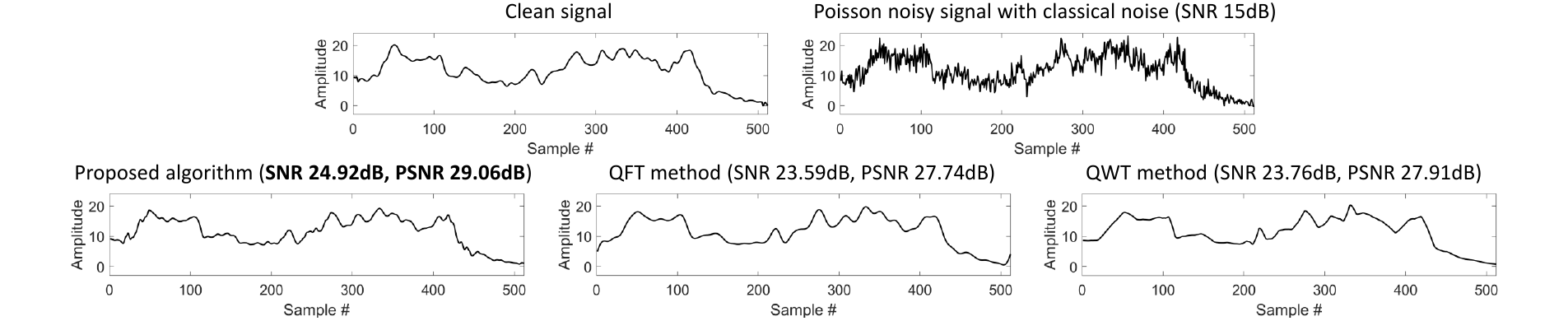}}
\vspace{-1mm}
\subfigure[Denoising performance in the presence of mixed noise (classical Poisson + quantum phase noise).]
{\includegraphics[width=1\textwidth]{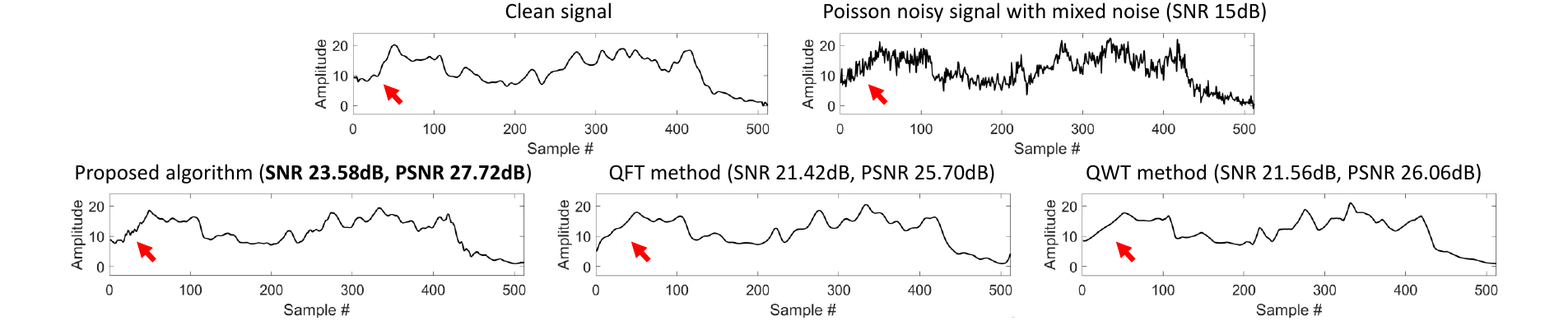}}

\caption{Denoising potential of different methods in various noise scenarios in presence of Poisson noise. The best results are highlighted in bold. The red arrow shows the reconstruction distortion arising in presence of quantum noise while using QFT and QWT, where the proposed method efficiently preserves the shape.}
\label{fig:denoi_result_poiss_supp}
\end{figure*}

\begin{figure*}[h!]
\vspace{-5mm}
\centering



\includegraphics[width=1\textwidth]{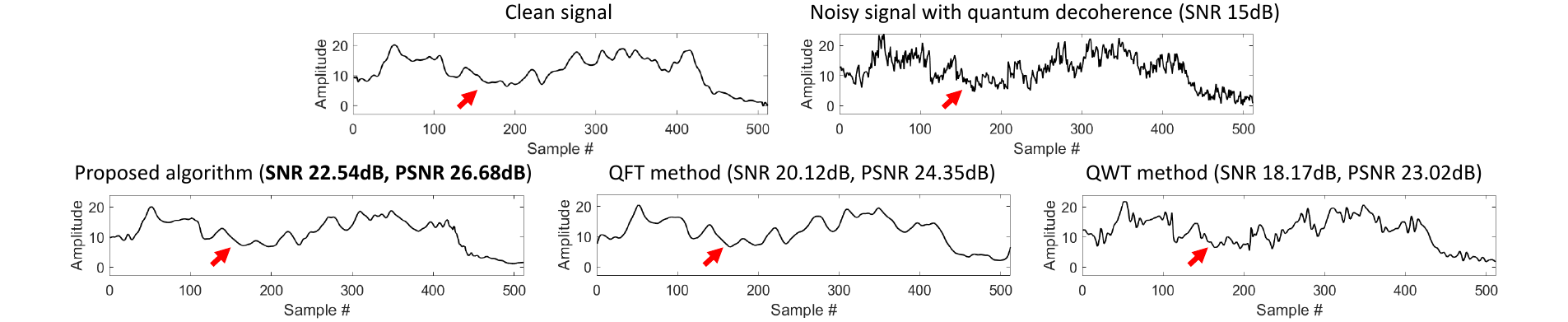}

\vspace{-1mm}
\caption{Denoising potential of different methods in presence of bit-flip noise and quantum phase noise. The best results are highlighted in bold. The red arrow shows the reconstruction distortion arising in presence of bit-flip noise while using QFT and QWT, where the proposed method preserves the shape more efficiently.}
\label{fig:decoh_result_supp}
\vspace{-2mm}
\end{figure*}

In this Supplementary Material, we define precisely the proposed unitary operation $\mcalbT$ to encode a classical signal into a quantum state, and present additional signal denoising results in presence of signal dependent Poissonian noise and mixed noise in Fig.~\ref{fig:denoi_result_poiss_supp}, as well as analyze the effect of bit-flip errors which appear in quantum decoherence processes in Fig.~\ref{fig:decoh_result_supp} using our proposed quantum algorithm and compared with standard Quantum Fourier Transforms (QFT) and Quantum Wavelet Transforms (QWT).
\vspace{-1mm}
\begin{footnotesize}
\begin{align}
\mcalbT = \prod_{j = 0}^{P-1} \Bigg( \bsI^{\bigotimes (m+a+1)} \bigotimes \Big( \sum_{l=0, l \neq j}^{P-1} \ket{l}^{\mbox{\scriptsize{J}}} \bra{l}^{\mbox{\scriptsize{J}}} \Big) \bigotimes \bsI + \prod_{i = 0}^{M-1}  \bigg(  \Big( \bsI \bigotimes \sum_{r=0, r \neq i}^{M-1} \ket{r}^{\mbox{\scriptsize{I}}}  \bra{r}^{\mbox{\scriptsize{I}}}  + \mcalbR_y(\theta_{ji}) \bigotimes \ket{i}^{\mbox{\scriptsize{I}}} \bra{i}^{\mbox{\scriptsize{I}}} \Big) \bigotimes  \mbfOmega_j \bigotimes \ket{j}^{\mbox{\scriptsize{J}}} \bra{j}^{\mbox{\scriptsize{J}}} \bigotimes \bsC_{\mbfOmega_j}  \bigg) \Bigg),\nonumber
\label{eq:uni_tran_T}
\end{align}
\end{footnotesize}
where $\bigotimes$ represents the tensor product, $\bsI $ is the identity transform, $\mcalbR_y(\theta_{ji})$ are the controlled rotation operations about the $y$-axis with angle $\theta_{ji}$, and, $\mbfOmega_j$ and $\bsC_{\mbfOmega_j}$ are the controlled value setting operations.

To implement quantum decoherence noises, we performed bit-flip operations randomly after each 1-qubit Hadamard gate in the quantum Fourier and inverse Fourier transforms together with quantum phase noise (by rotating each of the basic unitary operations with a small random angle of amplitude $\epsilon = 10^{-1}$, as explained in the main text Sec.IV).
Fig.~\ref{fig:decoh_result_supp} exemplifies the robustness of our proposed method against quantum decoherence noises.
On average, our proposed method produces denoised signals with 19.77 dB PSNR, whereas standard QFT and QWT algorithms yield 17.98 dB and 17.46 dB PSNRs, respectively. Therefore, the proposed method significantly outperforms previously proposed algorithms.

\end{document}